\documentclass[twocolumn,english,prl,showpacs,floatfix]{revtex4}
\usepackage{amsmath}
\usepackage{amsfonts}
\usepackage[T1]{fontenc}
\usepackage[latin1]{inputenc}
\usepackage{graphicx}
\usepackage{amssymb}
\usepackage{graphicx}
\usepackage{babel}



\begin{document}

\title{Atom-atom correlations and relative number squeezing in dissociation
of spatially inhomogeneous molecular condensates}
\author{Magnus \"{O}gren and K. V. Kheruntsyan}
\affiliation{ARC Centre of Excellence for Quantum-Atom Optics,
School of Physical Sciences, University of Queensland, Brisbane,
Queensland 4072, Australia}
\date{\today{}}

\begin{abstract}
We study atom-atom correlations and relative number squeezing in the
dissociation of a Bose-Einstein condensate (BEC) of molecular dimers made of
either bosonic or fermionic atom pairs. Our treatment addresses the role of
the spatial inhomogeneity of the molecular BEC on the strength of
correlations in the short time limit. We obtain explicit analytic results
for the density-density correlation functions in momentum space, and show
that the correlation widths and the degree of relative number squeezing are
determined merely by the shape of the molecular condensate.
\end{abstract}

\maketitle

Dissociation of a Bose-Einstein condensate (BEC) of molecular dimers \cite%
{Dissociation-exp} into pair-correlated atoms represents the
matter-wave analog of two-photon parametric down-conversion. The
latter process has been of crucial importance to the development of
quantum optics. Owing to this analogy, molecular dissociation
currently represents one of the \textquotedblleft
workhorses\textquotedblright\ of the new field of
quantum-atom optics \cite%
{Moelmer2001-twinbeams,Diss-others,Yurovski-diss,EPR,Fermidiss,Savage,SavageKheruntsyanSpatial,Jack-Pu,Diss-recent-others}
and offers promising opportunities for the generation of strongly correlated
atomic ensembles and fundamental tests of quantum mechanics with macroscopic
numbers of massive particles. Examples include the demonstration of the
Einstein-Podolski-Rosen paradox and violation of a classical Cauchy-Schwartz
inequality \cite{EPR,Kurizki,SavageKheruntsyanSpatial}. A closely related
process is atomic four-wave mixing in a collision of two BECs \cite%
{FWM-exp,Perrin-BEC-collisions}, which produces a spherical halo of
spontaneously scattered atoms \cite%
{Meystre-spin-EPR,Duan-spin-EPR,Yurovski-FWM,Trippenbach,NorrieBallaghGardiner-DeuarDrummond,Moelmer-width,Perrin-theory}
with correlations very similar to those in dissociation.

A recently emerged discussion topic -- following the experiments of Ref.~%
\cite{Perrin-BEC-collisions} on BEC collisions -- is the understanding of
the width and the strength of the observed correlations, as well as the
prospects of detecting relative number squeezing in the halo of the $s$-wave
scattered atoms \cite{Moelmer-width,Trippenbach,Perrin-theory}. The same
questions are relevant to atom-atom correlations in molecular dissociation
and have not been fully addressed so far.

In this paper, we study atom-atom correlations and relative number
squeezing in the dissociation of a molecular BEC in the short time
limit. Our analysis applies to molecules that may consist of pairs
of either bosonic or fermionic atoms, and takes into account the
spatial inhomogeneity of the
molecular BEC. It has been argued before and shown in the bosonic case \cite%
{Savage,SavageKheruntsyanSpatial} that the treatment of spatial
inhomogeneity is crucial for obtaining quantitatively correct results for
atom-atom correlations. In the fermionic case, the treatments of
dissociation have been so far restricted only to uniform systems \cite%
{Fermidiss,Jack-Pu,Corney-fermionic,PMFT}. In all cases, however, the
techniques are numerical and do not give the transparency of analytic
understanding, in contrast to the results obtained here.

The effective quantum field theory Hamiltonian describing our system in the
undepleted molecular field approximation is given, in a rotating frame, by
\cite{PRA2000-Feshbach-KKPD}%
\begin{align}
\widehat{H}& =\int dx\left\{ \sum\nolimits_{i=1,2}\left( \frac{\hbar ^{2}}{2m%
}|\mathbf{\partial }_{x}\widehat{\Psi }_{i}|^{2}+\hbar \Delta \widehat{\Psi }%
_{i}^{\dagger }\widehat{\Psi }_{i}\right) \right.  \notag \\
& \left. -i\hbar g(x)\left( \widehat{\Psi }_{1}\widehat{\Psi }_{2}-\widehat{%
\Psi }_{2}^{\dagger }\widehat{\Psi }_{1}^{\dagger }\right) \right\} .
\label{hamiltonian}
\end{align}%
The operators $\widehat{\Psi }_{1,2}(x,t)$ describe the atoms in two
different spin states, which can be either bosonic or fermionic, and we
assume that they have the same mass. The effective coupling $g(x)$ is defined as $%
g(x)=\chi \sqrt{\rho _{0}(x)}$, where $\chi $ is the atom-molecule coupling
(see \cite{PMFT} for details) and $\rho _{0}(x)$ is the initial density of
the molecular BEC in a harmonic trap. For computational simplicity we start
by treating a one-dimensional (1D) system; the analytic results will later
be generalized to three dimensions (3D).

The key difference between the present and previous (uniform) treatments of
dissociation in the undepleted molecular approximation \cite%
{Fermidiss,Yurovski-diss} is that we retain the spatial dependency of the
molecular BEC: the effective coupling $g(x)$ absorbs the molecular field,
which is treated classically via the coherent mean-field amplitude $\Psi
_{0}(x)=\sqrt{\rho _{0}\left( x\right) }$. The undepleted molecular
approximation is valid only for short dissociation times, during which the
converted fraction of molecules does not exceed $\sim 10\%$ \cite%
{Savage,PMFT}. Accordingly, the coupling $g(x)$ can be kept constant
in time, although the evolution of the atomic field is taking place
in free space. In this regime, the dissociation typically creates
low-density atomic clouds for which the $s$-wave scattering
interactions are a negligible effect too \cite{Savage}.

The detuning $\Delta $ in Eq.~(\ref{hamiltonian}) corresponds to the energy
mismatch $2\hbar \Delta $ between the free two-atom state in the
dissociation threshold and the bound molecular state. Molecules that are
unstable against spontaneous dissociation correspond to $\Delta <0$, with $%
2\hbar |\Delta |$ being the total dissociation energy that is converted into
kinetic energy of atom pairs primarily populating the resonant momenta
around $\pm k_{0}$, with $k_{0}=\sqrt{2m|\Delta |/\hbar }$.

Writing down the Heisenberg equations of motion for the field operators and
converting to Fourier space $\widehat{\Psi }_{j}(x,t)=\int dk\widehat{a}%
_{j}(k,t)\exp (ikx)/\sqrt{2\pi }$, we arrive at the following coupled
equations for the operators $\widehat{a}_{j}(k,t)$:%
\begin{gather}
\frac{d\widehat{a}_{1}(k,t)}{dt}=-i\Delta _{k}\widehat{a}_{1}(k,t)\pm \int
\frac{dq}{\sqrt{2\pi }}\widetilde{g}(q+k)\widehat{a}_{2}^{\dagger }(q,t),
\notag \\
\frac{d\widehat{a}_{2}^{\dagger }(k,t)}{dt}=i\Delta _{k}\widehat{a}%
_{2}^{\dagger }(k,t)+\int \frac{dq}{\sqrt{2\pi }}\widetilde{g}(q-k)\widehat{a%
}_{1}(-q,t).  \label{HeisenbergsEquation}
\end{gather}%
Here and hereafter the $+$ ($-$) [in general, upper (lower)] sign
stands for bosonic (fermionic) statistics of the atoms,
$\widetilde{g}(k)=\int dxe^{-ikx}g(x)/\sqrt{2\pi }$ is the Fourier
transform of the effective coupling $g(x)$, and $\Delta _{k}\equiv
\hbar k^{2}/\left( 2m_{1}\right) +\Delta $.

Equations~(\ref{HeisenbergsEquation}) can be solved numerically using
standard methods of linear operator algebra. One can show that, for vacuum
initial conditions, the only nonzero second-order moments are the normal and
anomalous atomic densities, $n_{j}(k,k^{\prime },t)\equiv \langle \widehat{a}%
_{j}^{\dagger }(k,t)\widehat{a}_{j}(k^{\prime },t)\rangle $ and $%
m_{12}(k,k^{\prime },t)\equiv \langle \widehat{a}_{1}(k,t)\widehat{a}%
_{2}(k^{\prime },t)\rangle $. From Eqs.~(\ref{HeisenbergsEquation}) we can
see that the finite width of $\widetilde{g}(k)$ -- due to the inhomogeneity
of the source -- implies that $\widehat{a}_{1}(k)$ couples not only to the
partner spin component at exactly opposite momentum $\widehat{a}%
_{2}^{\dagger }(-k)$ (as is the case in the homogeneous system), but also to
a range of momenta around $-k$, within $-k\pm \delta k$. The spread in $%
\delta k$ determines the width of atom-atom correlations and is ultimately
related to the width of the molecular BEC.

We now turn to the quantitative analysis of atom-atom correlations
expected to be present between the different spin-state atoms with
equal but opposite momenta due to momentum conservation, and between
the same spin-state atoms in the collinear direction due to quantum
statistical effects. We quantify these correlations via Glauber's
second-order correlation function,
\begin{equation}
g_{ij}^{(2)}(k,k^{\prime },t)=\frac{\langle \widehat{a}_{i}^{\dagger }(k,t)%
\widehat{a}_{j}^{\dagger }(k^{\prime },t)\widehat{a}_{j}(k^{\prime },t)%
\widehat{a}_{i}(k,t)\rangle }{n_{i}(k,t)n_{j}(k^{\prime },t)}.
\label{g2-def}
\end{equation}%
The normalization with respect to the product of densities $n_{i}(k,t)$ and $%
n_{j}(k^{\prime },t)$ [with $n_{j}(k,t)\equiv n_{j}(k,k,t)$] ensures that $%
g_{ij}^{(2)}(k,k^{\prime },t)=1$ for uncorrelated states. Due to obvious
symmetry considerations, $g_{12}^{(2)}(k,k^{\prime
},t)=g_{21}^{(2)}(k,k^{\prime },t)$ and $g_{11}^{(2)}(k,k^{\prime
},t)=g_{22}^{(2)}(k,k^{\prime },t)$.

Since the effective Hamiltonian corresponding to Eqs.~(\ref%
{HeisenbergsEquation}) is quadratic in the field operators, we can apply
Wick's theorem to factorize the fourth-order moment in Eq.~(\ref{g2-def}).
Noting that $\langle \widehat{a}_{1}^{\dagger }(k,t)\widehat{a}%
_{2}(k^{\prime },t)\rangle =\langle \widehat{a}_{j}(k,t)\widehat{a}%
_{j}(k^{\prime },t)\rangle =0$ in the present model, we obtain%
\begin{gather}
g_{12}^{(2)}(k,k^{\prime },t)=1+|m_{12}(k,k^{\prime
},t)|^{2}/[n_{1}(k,t)n_{2}(k^{\prime },t)],  \label{BB-corr-def} \\
g_{jj}^{(2)}(k,k^{\prime },t)=1\pm |n_{j}(k,k^{\prime
},t)|^{2}/[n_{j}(k,t)n_{j}(k^{\prime },t)].  \label{CL-corr-def}
\end{gather}

Before presenting the results based on numerical solutions of Eqs. (\ref%
{HeisenbergsEquation}), we now develop simple analytic approaches
that give approximate predictions for these observables, valid for
short times. More specifically, we treat the short time dynamics of
dissociation via the Taylor expansion in time, up to terms of order
$t^{2}$ \cite{comment1},
\begin{equation*}
\widehat{a}_{j}(k,t)=\widehat{a}_{j}(k,0)+\left. \frac{\partial \widehat{a}%
_{j}(k,t)}{\partial t}\right\vert _{t=0}t+\left. \frac{\partial ^{2}\widehat{%
a}_{j}(k,t)}{\partial t^{2}}\right\vert _{t=0}\frac{t^{2}}{2}+\ldots
\end{equation*}%
valid for $t\ll t_{0}$, where $t_{0}=1/%
\protect\chi \protect\sqrt{\protect\rho _{0}(0)}$ is the
time scale. Using the rhs of Eqs.~(%
\ref{HeisenbergsEquation}), this gives, up to the lowest-order terms, $%
n_{j}(k,k^{\prime },t)\simeq t^{2}\int dq\widetilde{g}^{\ast }(q+k)%
\widetilde{g}(q+k^{\prime })/2\pi $ and $|m_{12}(k,k^{\prime },t)|\simeq t|%
\widetilde{g}(k+k^{\prime })|/\sqrt{2\pi }$, or equivalently
\begin{gather}
n_{j}(k,k^{\prime },t)\simeq t^{2}\int dxe^{-i(k-k^{\prime
})x}[g(x)]^{2}/2\pi ,\;\;\;\;\;\;\;  \label{normal-moment-b} \\
|m_{12}(k,k^{\prime },t)|\simeq t\left\vert \int dxe^{-i(k+k^{\prime
})x}g(x)/2\pi \right\vert .  \label{anomalous-moment-b}
\end{gather}%
These results show that the width of the collinear (CL) correlation
between the same-spin atoms with nearly the same momenta,
Eq.~(\ref{CL-corr-def}), will be determined by the square of the
Fourier transform of the square of the effective coupling $g(x)$. On
the other hand, the width of the back-to-back (BB) correlation,
Eq.~(\ref{BB-corr-def}), between the different spin-state atoms with
nearly opposite momenta will be determined by the square of the
Fourier transform of $g(x)$. Therefore, the CL correlation is
generally broader than the BB correlation. These conclusions are
true for any shape of the source and apply to both bosonic and
fermionic statistics in the short time limit.

\textit{Thomas-Fermi (TF) parabolic density profile.} -- We now give
explicit analytic results for the case of a TF inverted parabola for
the molecular BEC density, $\rho _{0}(x)=\rho
_{0}(1-x^{2}/R_{\mathrm{TF}}^{2})$
($|x|\leq R_{\mathrm{TF}}$), in which case $g(x)=\chi \sqrt{\rho _{0}}%
(1-x^{2}/R_{\mathrm{TF}}^{2})^{1/2}$. Using the integral representation of
Bessel functions $J_{\nu }(z)$ \cite{MathHandbook}, Eqs.~(\ref%
{normal-moment-b}) and (\ref{anomalous-moment-b}) yield%
\begin{gather}
n_{j}(k,k^{\prime },t)\simeq \frac{2t^{2}\chi ^{2}\rho _{0}R_{TF}}{\sqrt{%
2\pi }}\;\frac{J_{3/2}\left( (k-k^{\prime })R_{TF}\right) }{\left[
(k-k^{\prime })R_{TF}\right] ^{3/2}},  \label{normal-Bessel} \\
|m_{12}(k,k^{\prime },t)|\simeq \frac{t\chi \sqrt{\rho _{0}}R_{TF}}{2}\;%
\frac{J_{1}\left( (k+k^{\prime })R_{TF}\right) }{(k+k^{\prime
})R_{TF}}. \label{anomalous-Bessel}
\end{gather}%
Since $J_{\nu }(z)\simeq (z/2)^{\nu }/\Gamma (\nu +1)$ for $z\ll 1$, the
atomic momentum distribution $n_{j}(k,t)$ and the diagonal anomalous density
$m_{12}(k,-k,t)$ are $n_{j}(k,t)\simeq 2t^{2}\chi ^{2}\rho _{0}R_{\mathrm{TF%
}}/3\pi $ and $|m_{12}(k,-k,t)|\simeq t\chi \sqrt{\rho _{0}}R_{\mathrm{TF}%
}/4 $. Despite the fact that the atomic momentum distribution in the lowest order in $%
t$ is uniform, the momentum cutoff $k_{\max }$ \cite{PRA2000-Feshbach-KKPD}
-- which must be assumed when using a $\delta$-function interaction in Eq.~(\ref%
{hamiltonian}) -- prevents the total atom number from diverging.

Substituting Eqs.~(\ref{normal-Bessel}) and~(\ref{anomalous-Bessel}) into Eqs.~(%
\ref{BB-corr-def}) and~(\ref{CL-corr-def}), we obtain the following explicit
results for the atom-atom correlations, valid for $t\ll t_{0}$:
\begin{gather}
g_{12}^{(2)}(k,k^{\prime },t)\simeq 1+\frac{9\pi ^{2}}{16t^{2}\chi
^{2}\rho _{0}}\frac{\left[ J_{1}\left( (k+k^{\prime
})R_{\mathrm{TF}}\right) \right]
^{2}}{\left[ (k+k^{\prime })R_{\mathrm{TF}}\right] ^{2}},  \label{g12} \\
g_{jj}^{(2)}(k,k^{\prime },t)\simeq 1\pm \frac{9\pi }{2}\frac{\left[ J_{3/2}%
\left( (k-k^{\prime })R_{\mathrm{TF}}\right) \right] ^{2}}{\left[
(k-k^{\prime })R_{\mathrm{TF}}\right] ^{3}}.  \label{gjj}
\end{gather}

\begin{figure}[tbp]
\includegraphics[height=3.8cm]{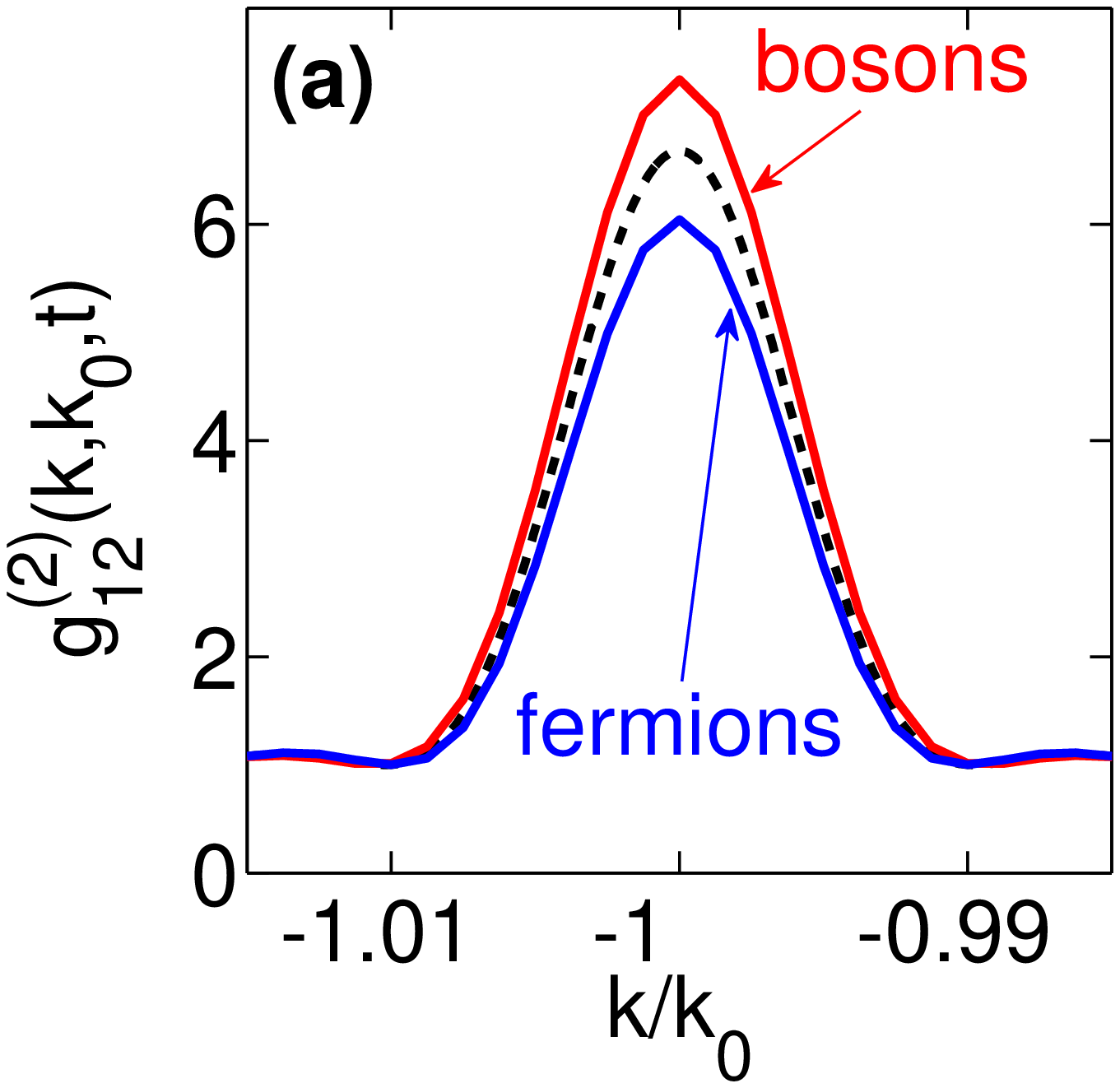} %
\includegraphics[height=3.8cm]{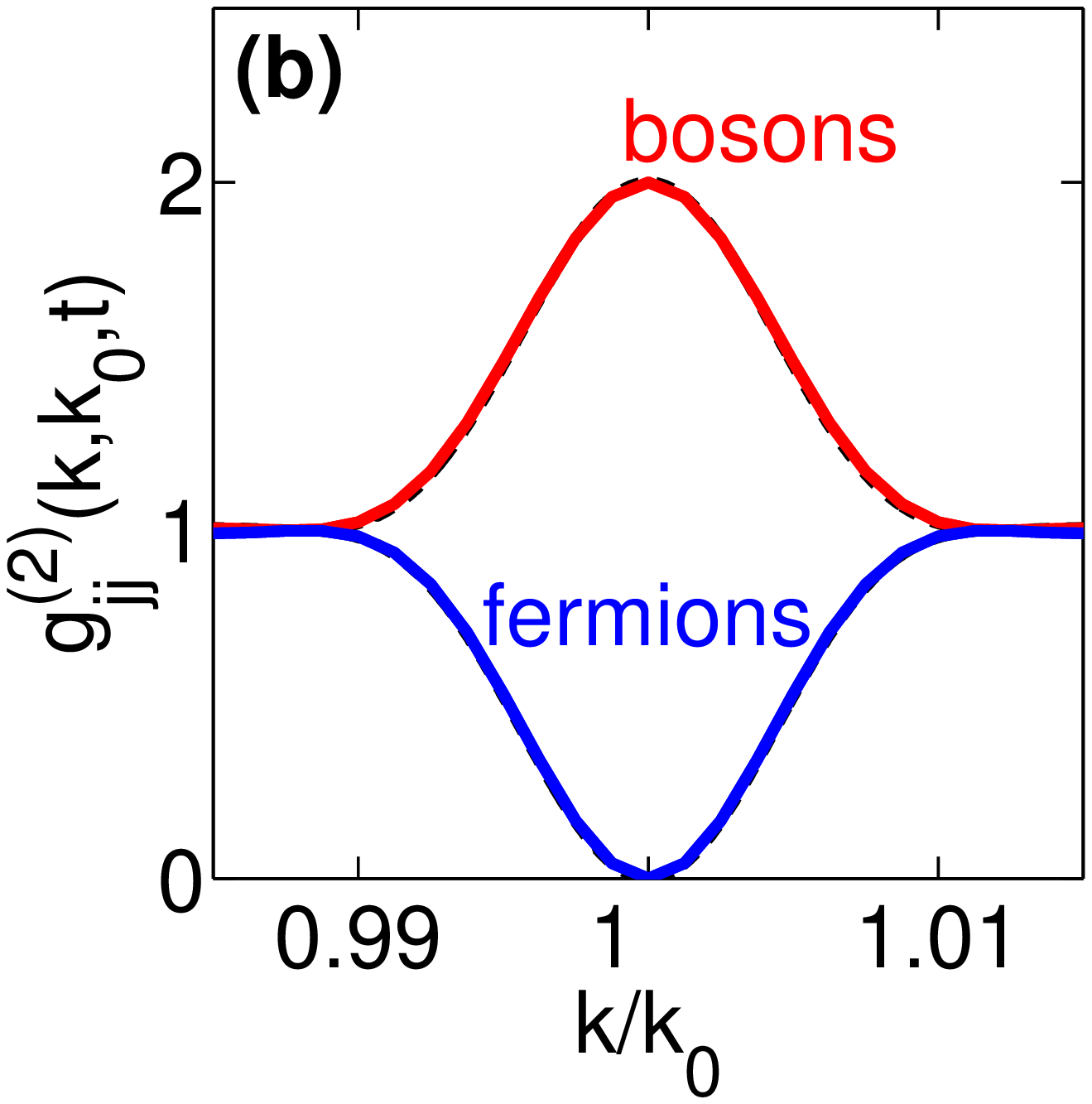}
\caption{Back-to-back (a) and collinear (b) correlation $%
g_{ij}^{(2)}(k,k_{0},t)$ as a function of $k$ at $t/t_{0}=0.5$. The
dimensionless detuning is $\protect\delta =\Delta t_{0}=-9$, where $t_{0}=1/%
\protect\chi \protect\sqrt{\protect\rho _{0}(0)}\simeq 5$ ms is the time
scale; the Thomas-Fermi radius of the molecular BEC is $R_{\mathrm{TF}}=250$
$\protect\mu $m; for other parameters, see Ref. \protect\cite%
{PhysicalParameters}.}
\label{g2-a-b-c}
\end{figure}

The pair correlations $g_{ij}^{(2)}(k,k_{0},t=0.5t_{0})$, where the
momentum of one of the atomic components is fixed to $k_{0}$, while
the partner momentum $k$ is varied, is plotted in
Figs.~\ref{g2-a-b-c}(a) and~\ref{g2-a-b-c}(b). The dashed lines are
the analytic results of Eqs.~(\ref{g12}) and (\ref{gjj}), whereas
the solid lines are the numerical results from Eqs.~(\ref%
{HeisenbergsEquation}). In Fig.~\ref{g2-a-b-c}(b) the dashed lines
are almost indistinguishable from the respective solid lines, even
though the ratio $t/t_{0}=0.5$ is not very small. For earlier times,
the agreement between the analytic and numerical results is even
better.

In the case of different spin states, $g_{12}^{(2)}(k,k_{0},t)$,
with either bosonic or fermionic atoms, we see a strong BB
correlation
signal between atom pairs with equal but opposite momenta, centered at $%
k=-k_{0}$. The same-spin CL correlation function
$g_{jj}^{(2)}(k,k_{0},t)$, on the other hand, shows the Hanbury
Brown and Twiss bunching peak for bosons
and an antibunching dip for fermions due to Pauli blocking \cite%
{Fermi-antibunching}, both centered at $%
k=k_{0}$.

\begin{figure}[tbp]
\includegraphics[height=3.8cm]{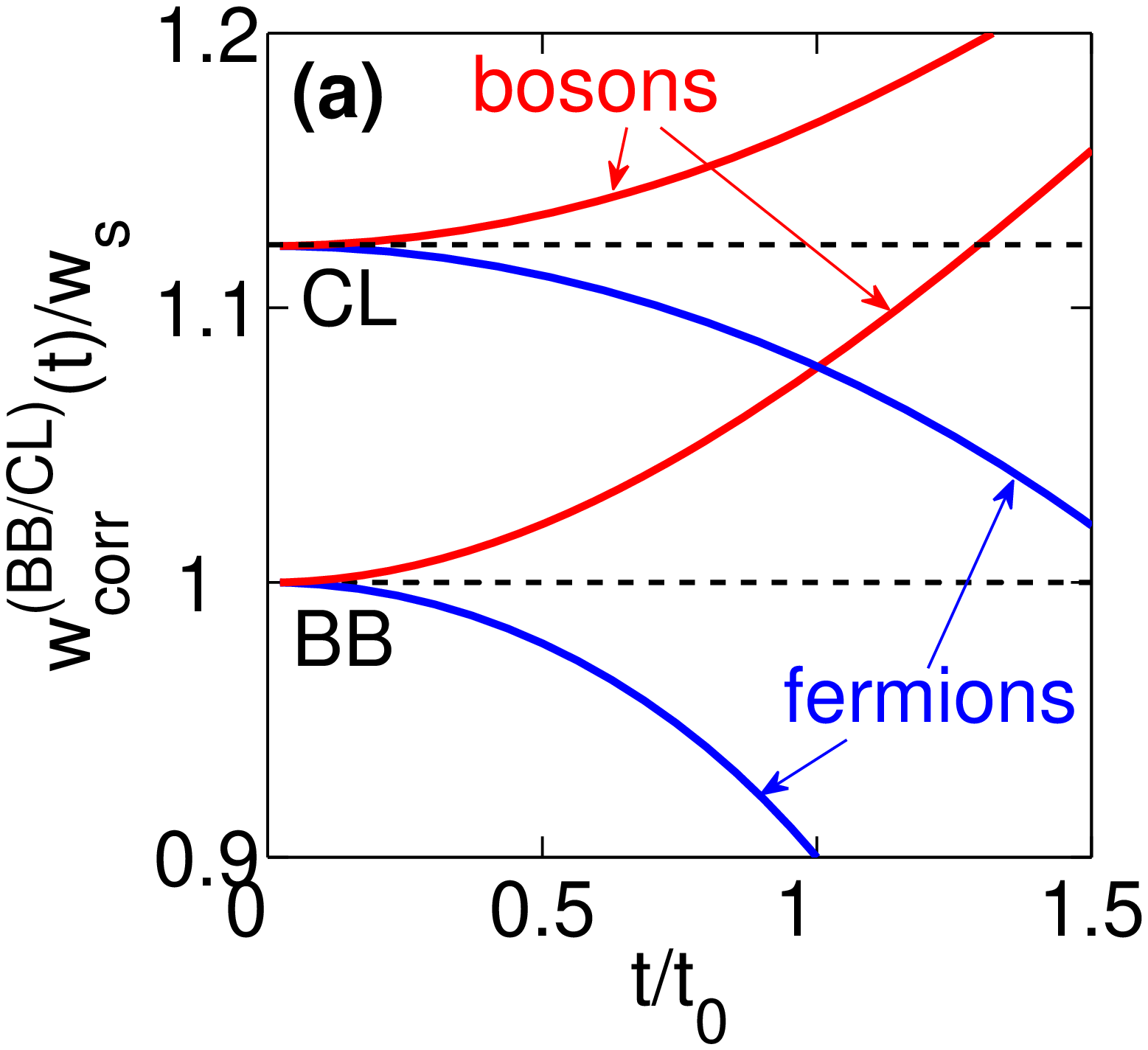} %
\includegraphics[height=3.8cm]{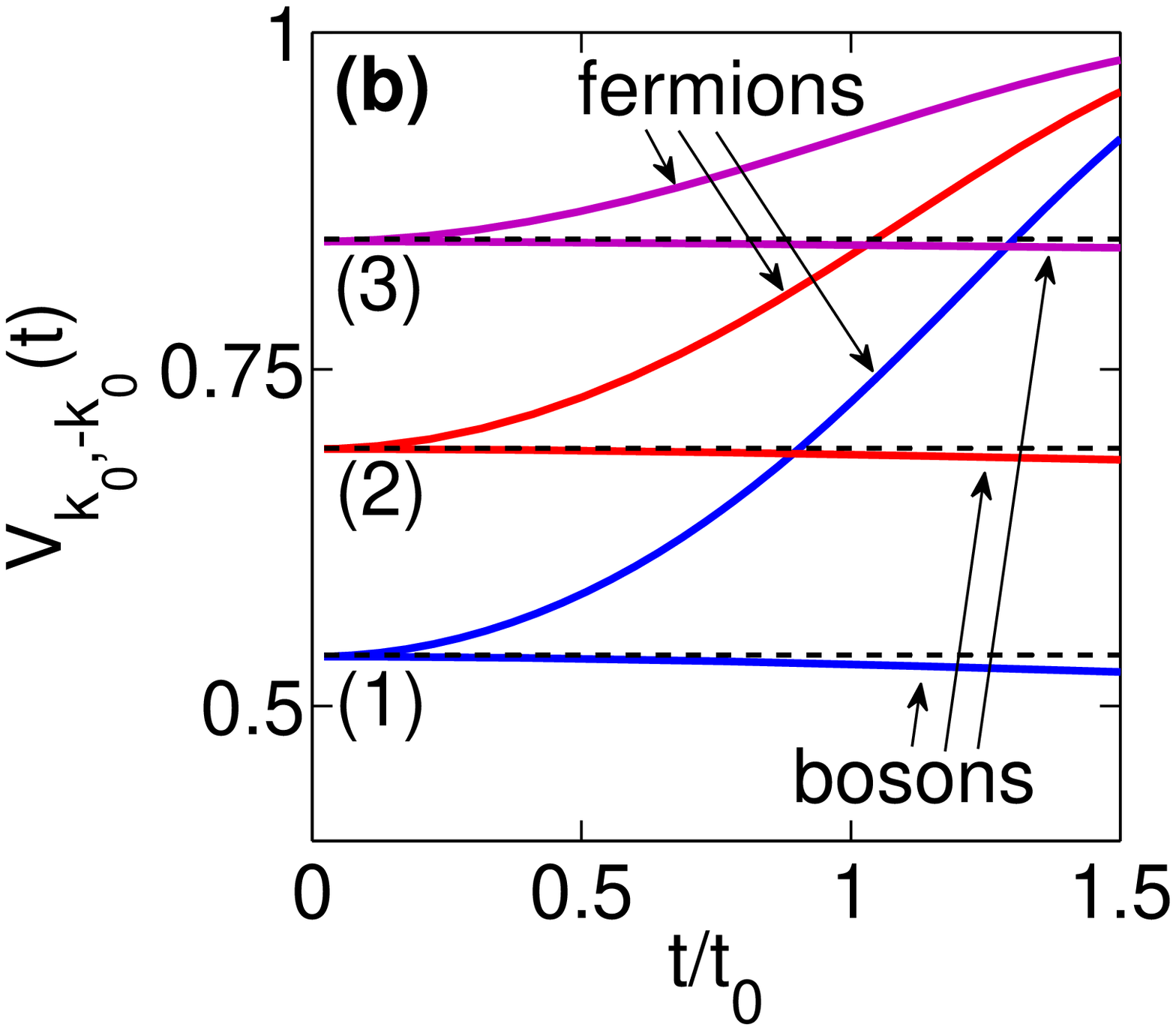}
\caption{(a) Width of the BB and CL correlations relative to the momentum
width of the molecular BEC, $w_{s}\simeq 1.62/R_{\mathrm{TF}}$, as a
function of time, for the physical parameters of Fig. \protect\ref%
{g2-a-b-c}. (b) Relative number variance $V_{k_{0},-k_{0}}(t)$ as a function
of time, for $R_{\mathrm{TF}}^{(1)}=250$ $\protect\mu $m (1), $R_{\mathrm{TF}%
}^{(2)}=167$ $\protect\mu $m (2), and $R_{\mathrm{TF}}^{(3)}=83$ $\protect%
\mu $m (3). The counting length is $\Delta k=\protect\pi /2R_{\mathrm{TF}%
}^{(1)}$ in all cases. }
\label{widths-a-b-c-d}
\end{figure}

In Fig.~\ref{widths-a-b-c-d}(a), we plot the widths of the BB and CL
correlations as a function of time. For simplicity, we define them
as the half-width at half-maximum. The widths in the bosonic and
fermionic cases (with the solid lines corresponding to the numerical
results) have universal asymptotics in the limit $t\rightarrow 0$,
when the quantum statistical effects are irrelevant due to low mode
occupancies. The asymptotic values
(dashed lines) are found from Eqs. (\ref{g12}) and (\ref{gjj}): $w_{\mathrm{%
corr}}^{(\mathrm{BB})}=w_{\mathrm{s}}$ and $w_{\mathrm{corr}}^{(\mathrm{CL}%
)}\simeq 1.12w_{\mathrm{s}}$, where $w_{\mathrm{s}}\simeq 1.62/R_{\mathrm{TF}%
}$ is the width of the momentum distribution $\widetilde{\rho }_{0}(k)=$ $|%
\widetilde{g}(k)|^{2}/\chi ^{2}=|\sqrt{\rho _{0}\pi /2}J_{1}(kR_{\mathrm{TF}%
})/k|^{2}$ of the actual source -- the molecular BEC.

As an alternative measure of the strength of atom-atom correlations \cite%
{Savage}, we also calculate the variance of number-difference
fluctuations for atoms in different spin states and with equal but
opposite momenta $\pm k_{0}$,
\begin{equation}
V_{k_{0},-k_{0}}(t)=\langle \lbrack \Delta (\widehat{N}_{1,k_{0}}-\widehat{N}%
_{2,-k_{0}})]^{2}\rangle /SN,  \label{Variance-def}
\end{equation}%
where $SN$ is the shot-noise level that originates from uncorrelated states.
The number operators are defined by $\widehat{N}_{j,\pm k_{0}}(t)=\int_{K}dk%
\widehat{n}_{j}(k,t)$ [with $\widehat{n}_{j}(k,t)=\widehat{a}_{j}^{\dagger
}(k,t)\widehat{a}_{j}(k,t)$], where $K$ is the counting length around $\pm
k_{0}$. On a computational lattice the simplest choice that does not require
explicit binning of the signal is $K=$ $\Delta k$, where $\Delta k$ is the
lattice spacing, and therefore $\widehat{N}_{j,\pm k_{0}}(t)=\widehat{n}%
_{j}(\pm k_{0},t)\Delta k$. We emphasize that $SN$ is different for
bosons and fermions. For the bosonic case, $SN$ is given by the sum
of variances of the individual mode occupancies with Poissonian
statistics (as in the
coherent state), implying that $SN=\langle \widehat{N}_{1,k_{0}}\rangle +$ $%
\langle \widehat{N}_{2,-k_{0}}\rangle $. For the fermionic case, the sum of
individual variances gives $SN=\langle \widehat{N}_{1,k_{0}}\rangle
(1-\langle \widehat{N}_{1,k_{0}}\rangle )+$ $\langle \widehat{N}%
_{2,-k_{0}}\rangle (1-\langle \widehat{N}_{2,-k_{0}}\rangle )$ \cite%
{Fermidiss}. The variance (\ref{Variance-def}) can be rewritten as
\begin{eqnarray}
V_{k_{0},-k_{0}}(t) &=&1-\frac{\Delta kn_{1}(k_{0},t)}{1-s\Delta
kn_{1}(k_{0},t)}  \notag \\
&&\times \lbrack g_{12}^{(2)}(k_{0},-k_{0},t)-g_{11}^{(2)}(k_{0},k_{0},t)],
\label{V-raw}
\end{eqnarray}%
where $s=0(1)$ for bosons (fermions), and we have taken into account that $%
\langle \widehat{N}_{1,k_{0}}\rangle =\langle \widehat{N}_{2,-k_{0}}\rangle $
and $g_{11}^{(2)}(k_{0},k_{0},t)=g_{22}^{(2)}(-k_{0},-k_{0},t)$. Variance $%
V_{k_{0},-k_{0}}(t)<1$ implies squeezing of fluctuations below the
shot-noise level and corresponds to a violation of the classical
Cauchy-Schwartz inequality with $%
g_{12}^{(2)}(k_{0},-k_{0},t)>g_{11}^{(2)}(k_{0},k_{0},t)$ \cite%
{SavageKheruntsyanSpatial}.

The short time asymptotics for the variance can be found using Eqs. (\ref%
{g12}) and (\ref{gjj}), yielding%
\begin{equation}
V_{k_{0},-k_{0}}(t\ll t_{0})=1-3\pi \Delta {}k\,R_{\mathrm{TF}}/32.
\label{V-TF-1D}
\end{equation}%
The small geometric prefactor in the second term, together with the
resolution requirement $\Delta k\lesssim 1/R_{\mathrm{TF}}$, ensures that $%
V_{k_{0},-k_{0}}>0$. We see that the squeezing is stronger for larger
condensates and counting lengths.

In Fig. \ref{widths-a-b-c-d}(b) we plot the variance $V_{k_{0},-k_{0}}(t)$
for three different sizes of the molecular condensate. The solid lines are
the numerical results from Eqs.~(\ref{HeisenbergsEquation}), whereas the
horizontal dashed lines are the short time asymptotic results of Eq.~(\ref%
{V-TF-1D}) matching precisely the numerical results in the limit $%
t\rightarrow 0$. As we can see, the squeezing of the relative number
fluctuations (from the numerical curves) does not change
significantly with time for bosons, while for fermions its dynamics
is affected by a stronger dependence of the fermionic shot noise on
the mode occupancy. The squeezing is stronger
for larger condensates, but is still far from perfect squeezing, $%
V_{k_{0},-k_{0}}(t)=0$, which follows from the idealized uniform models \cite%
{Savage,Fermidiss}.

\textit{Gaussian density profile.} -- For comparison, we also give
the analytic results for a Gaussian density profile $\rho
_{0}(x)=\rho _{0}\exp (-x^{2}/2\sigma _{x}^{2})$ of the molecular
BEC, giving the momentum distribution of $\widetilde{\rho
}_{0}(k)\propto \exp (-k^{2}/2\sigma _{k}^{2})$, where $\sigma _{x}$
and $\sigma _{k}=1/2\sigma _{x}$ are the rms
widths. In this case, the atom-atom correlations are%
\begin{gather}
g_{12}^{(2)}(k,k^{\prime },t\ll t_{0})\simeq 1+2e^{-(k+k^{\prime
})^{2}/2\sigma _{k}^{2}}/(t^{2}\chi ^{2}\rho _{0}), \\
g_{jj}^{(2)}(k,k^{\prime },t\ll t_{0})\simeq 1\pm e^{-(k-k^{\prime
})^{2}/4\sigma _{k}^{2}}.
\end{gather}%
The respective rms widths are given by $\sigma _{\mathrm{corr}}^{(\mathrm{BB}%
)}=\sigma _{k}$ and $\sigma _{\mathrm{corr}}^{(\mathrm{CL})}=\sqrt{2}\sigma
_{k}$, resulting in the ratio $\sigma _{\mathrm{corr}}^{(\mathrm{CL}%
)}/\sigma _{\mathrm{corr}}^{(\mathrm{BB})}=\sqrt{2}$. The relative number
squeezing is $V_{k_{0},-k_{0}}(t\ll t_{0})=1-\sqrt{2/\pi }\,\Delta k\,\sigma
_{x}$, with $\Delta k\lesssim 1/2\sigma _{x}$ ensuring $V_{k_{0},-k_{0}}>0$.

\textit{Results in 3D.} -- For a nonisotropic TF parabolic density
profile
of the molecular BEC, performing the integrals in 3D versions of Eqs.~(\ref%
{normal-moment-b})~and~(\ref{anomalous-moment-b}) is more cumbersome and we
only give the final results for correlations corresponding to the
displacement between the pairs of momenta along one of the Cartesian
coordinates $\alpha =x,y,z$. The BB and CL correlation widths obtained here
are $w_{\mathrm{corr,}\alpha }^{(\mathrm{BB})}=w_{\mathrm{s,}\alpha }$ and $%
w_{\mathrm{corr,}\alpha }^{(\mathrm{CL})}\simeq 1.08w_{\mathrm{s,}\alpha }$,
where $w_{\mathrm{s,}\alpha }\simeq 1.99/R_{\mathrm{TF,}\alpha }$. The
relative number variance is $V_{k_{0},-k_{0}}(t\ll t_{0})\simeq 1-15(%
\overline{\Delta k}\;\overline{R_{\mathrm{TF}}})^{3}/2^{10}$, where $%
\overline{\Delta k}=(\Delta k_{x}\Delta k_{y}\Delta k_{z})^{1/3}$ and $%
\overline{R_{\mathrm{TF}}}=(R_{\mathrm{TF,}x}R_{\mathrm{TF,}y}R_{\mathrm{TF,}%
z})^{1/3}$ are the geometric means. The much smaller (than in 1D) geometric
prefactor in the second term, together with $\Delta k_{\alpha }\lesssim 1/R_{%
\mathrm{TF,}\alpha }$, explains why the \textquotedblleft
raw\textquotedblright\ (unbinned) squeezing is much weaker in 3D~\cite%
{Savage,SavageKheruntsyanSpatial} than in 1D. Therefore, the
prescription of Ref.~\cite{SavageKheruntsyanSpatial} to perform
binning for obtaining stronger squeezing is more crucial in 3D.

For a Gaussian density profile, the generalization to 3D is straightforward.
In particular, the BB and CL correlation widths for a displacement along $x$%
, $y$, or $z$ are as in 1D, while the relative number variance is $%
V_{k_{0},-k_{0}}(t\ll t_{0})=1-[\sqrt{2/\pi }\;\;\overline{\Delta k}\,%
\overline{\sigma }]^{3}$, where $\overline{\sigma }=(\sigma _{x}\sigma
_{y}\sigma _{z})^{1/3}$.

In summary, we have studied the dissociation of a BEC of molecular dimers
into correlated fermionic and bosonic atom pairs. We have obtained explicit
analytic results for the width and strength of the atom-atom correlations
and for the relative number squeezing in the short time limit, using
realistic density profiles of the molecular BEC. The results show how the
squeezing improves with the larger size of the molecular condensate, and how
it can degrade in strongly inhomogeneous systems. Our approach can be easily
generalized to describe similar effects in atomic four-wave mixing via BEC
collisions \cite{Perrin-BEC-collisions,Moelmer-width,Perrin-theory}.

The authors acknowledge support from the Australian Research Council
and thank J. F. Corney, M. J. Davis, M. K. Olsen, and C. M. Savage
for useful discussions.


\begin{thebibliography}{99}
\bibitem{PhysicalParameters} In the case of fermionic atoms, we consider a
BEC of $^{40}$K$_{2}$ dimers with a peak 1D density $\rho
_{0}(0)=4\times 10^{7}$ m$^{-1}$ and $\chi =3.15\times 10^{-2}$
m$^{1/2}$/s \cite{PMFT}. In the bosonic case, we assume the same
parameters as a hypothetic example.
\bibitem{comment1} One can check that the commutation/anticommutation
relations here are $\left[ \widehat{a}(\mathbf{k},t),\widehat{a}^{\dagger }(%
\mathbf{k}^{\prime },t)\right]_{\mp} \approx \delta (\mathbf{k}-\mathbf{k}%
^{\prime })$ up to terms of the order of $t^{2}$.
\end{thebibliography}
\end{document}